\def\papertitle{RTNeural: Fast Neural Inferencing for Real-Time Systems}
\def\paperauthorA{Jatin Chowdhury}

\documentclass[twoside,a4paper]{article}
\usepackage{dafx_19}
\usepackage{amsmath,amssymb,amsfonts,amsthm}
\usepackage{euscript}
\usepackage[latin1]{inputenc}
\usepackage[T1]{fontenc}
\usepackage{ifpdf}

\usepackage[english]{babel}
\usepackage{caption}
\usepackage{subfig} 
\usepackage{xcolor}
\definecolor{codegreen}{rgb}{0,0.6,0}

\usepackage{listings}
\ninept

\usepackage{times}

\newif\ifpdf
\ifx\pdfoutput\relax
\else
   \ifcase\pdfoutput
      \pdffalse
   \else
      \pdftrue
\fi

\ifpdf 
  \usepackage[pdftex,
    pdftitle={\papertitle},
    pdfauthor={\paperauthorA},
    colorlinks=false, 
    bookmarksnumbered, 
    pdfstartview=XYZ 
  ]{hyperref}
  \usepackage[pdftex]{graphicx}
  \usepackage[figure,table]{hypcap}
\else 
  \usepackage[dvips]{epsfig,graphicx}
  \usepackage[dvips,
    colorlinks=false, 
    bookmarksnumbered, 
    pdfstartview=XYZ 
  ]{hyperref}
  \usepackage[figure,table]{hypcap}
\fi

\usepackage{tikz}
\usetikzlibrary{positioning}
\usepackage[american]{circuitikz}
\usepackage{tkz-euclide}
\usepackage{cleveref}
\usepackage{siunitx}

\crefname{lstlisting}{listing}{listings}
\Crefname{lstlisting}{Listing}{Listings}

\DeclareMathAlphabet{\mathpzc}{OT1}{pzc}{m}{it}

\title{\papertitle}

\affiliation{
\paperauthorA \,}
{\href{http://ccrma.stanford.edu}{Center for Computer Research in Music and Acoustics} \\ Stanford University \\ Palo Alto, CA \\ {\tt \href{mailto:jatin@ccrma.stanford.edu}{jatin@ccrma.stanford.edu}}}

\begin{document}
\ifpdf 
  \DeclareGraphicsExtensions{.png,.jpg,.pdf}
\else  
  \DeclareGraphicsExtensions{.eps}
\fi

\graphicspath{{./Figures/}}
\def\RT{\texttt{RTNeural}}
\def\CPP{\texttt{C++}}
\def\footlink#1{\footnote{\href{#1}{#1}}}

\maketitle
\begin{abstract}
\RT\ is a neural inferencing library written in \CPP.
\RT\ is designed to be used in systems with hard real-time
constraints, with additional emphasis on speed, flexibility,
size, and convenience. The motivation and design of the
library are described, as well as real-world use-cases, and
performance comparisons with other neural inferencing libraries.
\end{abstract}

\section{Introduction}
Neural network models are typically trained using high-level
scripting languages, such as \texttt{Python}, often using
machine-learning packages including TensorFlow\footlink{https://github.com/tensorflow/tensorflow}
or PyTorch.\footlink{https://github.com/pytorch/pytorch}
Once training has completed, the neural network may be
used for ``inference''. While training utilises both
forward propagation to generate outputs, and backwards
propagation to update the network parameters, inferencing
uses only forward propagation to compute outputs for
a given set of inputs.
\newline\newline
For many real-world applications, it can be useful to implement
neural network inferencing in a more performant language such as
\CPP. To serve that purpose, TensorFlow and PyTorch contain
a \CPP\ API, which is often used to perform inference for mobile
and desktop applications that use neural network models. While these
libraries are flexible and efficient, especially for larger networks,
they are often inadequate for real-time systems.
\newline\newline
The goal of \RT\ is to maintain a ``real-time safe'' alternative
to the \CPP\ API's of TensorFlow and PyTorch.
Since certain classes of neural network architectures are
not practical for use in real-time systems, \RT\ only attempts
to implement a small subset of neural network building blocks compared
to other major libraries. While \RT\ is unable match the run-time
flexibility of these large libraries, it aims to be convenient
for programmers by providing several choices for computational backends
and options for defining the model architecture either at run-time
or compile-time.

\section{Real-Time Systems}
For the purposes of this writing, consider a generic
real-time system as a system which recieves blocks of input
data at a given frequency, and has a set amount of time in which
to process that data and compute any necessary outputs. A common
example of this type of system is real-time audio processing,
where a stream of audio samples at a given sample rate
are fed to the processor in blocks. If the processor is unable
to process the block fast enough to keep up with the system
sample rate, the audio stream will contain a block of ``incorrect''
samples, often resulting in an audible ``glitch''.

\subsection{Constraints of Real-Time Systems}
The fundamental reason why major neural network libraries
including PyTorch and TensorFlow are not adequate for real-time
use is that they do not strictly obey the rules of real-time programming.
While the rules of real-time programming are not well-defined
in scientific literature, a useful discussion is provided in
\cite{rt101}. The most generic rule of real-time programming
is to avoid calling code (within the real-time processing code)
that may take an ``unbounded'' amount of time, which includes
operations such as thread locking, memory allocation, ``garbage
collection'', and interfacing with external hardware.
\newline\newline
Unfortunately, most neural network libraries rely on
dynamically resizable data structures, such as \CPP's
\texttt{std::vector}, which require memory allocation
for most common operations. While this design choice
allows for more convenient and flexible implementations,
it slows down the real-time processing and leads to ``spikes''
in performance whenever the program needs to wait for the
operating system to allocate additional memory. While the
performance impact may be negligible for larger neural
networks, it can be significant for small neural networks,
which are often more practical for real-time applications.
\newline\newline
Note that while many neural networks used in production
run inference on a dedicated GPU to make better use of
their highly parallelised architecture, this approach
is poorly-suited for many real-time systems due to the overhead
of interfacing with the external hardware, although recent
progress has been made in this area \cite{travis_gpu}. However,
due to the fact that smaller neural networks are typically
better-suited for real-time inferencing, running inference
on a CPU that supports SIMD instructions is often as fast if
not faster than running on a GPU \cite{NLML}.

\section{Library Design}
\RT\ is designed as a lightweight, real-time safe
neural network inferencing library. To enable flexibility
for programmers using the library, \RT\ supports three computational
backends, as well as both run-time and compile-time APIs
for loading trained neural networks and performing inference.
\newline\newline
\RT\ currently supports the following neural network layers:
\begin{itemize}
    \item Fully-Connected (Dense)
    \item 1-Dimensional Convolution (Conv1D)
    \item Long Short-Term Memory (LSTM) \cite{lstm_original}
    \item Gated Recurrent Unit (GRU) \cite{gru_original}
\end{itemize}
As well as the following activation layers:
\begin{itemize}
    \item $tanh$
    \item Rectified Linear Unit (ReLU)
    \item Sigmoid
    \item SoftMax
\end{itemize}
\RT\ is open-source under the Berkeley Software Distribution (BSD)
3-clause license. The source code is publicly available on
GitHub.\footlink{https://github.com/jatinchowdhury18/RTNeural}

\subsection{Layer Implementations}
As mentioned above, not all layer types and layer sizes are
suitable for real-time systems. For example, when analyzing
the networks proposed for real-time use in \cite{NLML,VArnn,WaveNetVA,time_vary_rnn},
the only layer types used are Dense, Conv1D, LSTM, and GRU,
along with a handful of activation layers.
Further, none of the layers in the networks proposed requires
an input size larger than 100 channels; rather, the typical
input size for the proposed real-time neural network layers
is between 8-32 channels. With that in mind, \RT\ only
implements a subset of the many neural network layers that
may be used generally, and explicitly attempts to optimize
performance for smaller layer sizes.

\subsection{Computational Backends}
\RT\ supports three choices for the computational backend used
by the library: Eigen \cite{eigenweb}, XSIMD,\footlink{https://github.com/xtensor-stack/xsimd}
and the \CPP\ Standard Template Library (STL). There
is also experimental support for a fourth backend
using Apple's Accelerate framework,\footlink{https://developer.apple.com/documentation/accelerate}
however this backend can only be used on Apple devices,
and does not have complete support for all functionality
provided by the other backends.

\subsubsection{The Eigen Backend}
Eigen is a highly-templated matrix computation library, with
support for SIMD operations and vectorized maths functions.
As well as being the cleanest and most readable implementation,
the Eigen backend maintains good performance for networks
of all sizes, performing particularly well as the network grows larger.

\subsubsection{The XSIMD Backend}
XSIMD is a \CPP\ library which provides a common interface
for SIMD instruction sets across different processor architectures.
Therefore, the XSIMD backend implements the various neural network
layers essentially using direct SIMD instructions. These implementations
can avoid some of the overhead incurred by the Eigen implementations,
leading to better performance for small neural networks.

\subsubsection{The STL Backend}
Since some compilers (particularly for embedded devices)
may have difficulty compiling or linking with the Eigen
or XSIMD libraries, the STL backend is provided as a fallback.
Although the STL backend is typically the least performant, it can
easily be compiled with any modern \CPP\ compiler.
Note that since the choice of backends is made at compile-time,
the library will automatically exclude any unused libraries
from the build process.

\subsection{Loading Pre-Trained Models}
Since \RT\ does not support training neural network
models, the user must instead train the model themselves,
and then load the model into \RT. To make this process more
convenient, \RT\ provides \texttt{Python} scripts for saving
sequential models trained using the TensorFlow \texttt{Keras} API
to a JSON format, which can then be parsed by methods in \RT.
For non-sequential model architectures, or models trained using
a different framework, the user may export the model weights in
whatever form they choose, and then load the weights into an
\RT\ model, using the weight-setting functions provided with each layer.

\subsection{Run-Time API}
The \RT\ run-time API allows the user to load pre-trained
models at run-time and perform inference on those models.
Although the run-time API requires dynamic memory allocation
since the number of input and output channels for each layer
is not known at compile-time, \RT\ allocates all required memory
while loading the model so that memory is never allocated as
part of the real-time process. Note that while the run-time
API is more flexible than the compile-time API, it has somewhat
worse performance, since the compiler is unable to fully optimize
certain parts of the inferencing process that are not known
until the model is loaded at run-time. \Cref{lis:RunTimeEx}
shows a simple example of running inference for a neural
network with 3 input channels and 1 output channel, using
the run-time API.
\begin{lstlisting}[label={lis:RunTimeEx},captionpos=b,
    caption={\it Example of the RTNeural run-time API.
        Note that the type returned by \texttt{parseJson}
        is a \texttt{std}::\texttt{unique\_ptr<>}.
        }]
#include <RTNeural/RTNeural.h>
  
int main()
{
    // load model from json
    std::ifstream jsonStream("model.json", std::ifstream::binary);
    auto model = RTNeural::json_parser::parseJson<double>(jsonStream);
  
    model->reset(); // reset model state
  
    // set up input vector
    double input[] = { 1.0, 0.5, -0.1 };

    // compute output
    double output = model->forward(input);
  
    return 0; // exit
}
\end{lstlisting}
\subsection{Compile-Time API}
It is typically preferable to use the \RT\ compile-time API in
cases where the model architecture is known at compile-time.
With the compile-time API, all required memory is allocated
statically, and all functions needed for real-time inferencing
are defined such that they can be inlined by the compiler
\cite{inlining}. Further, since the number of input and output
channels needed for each layer is known at compile-time,
the compiler can ``unroll'' loops as needed to optimize
performance \cite{compiler_design}. \Cref{lis:CompileTimeEx}
shows a simple example of running inference for a neural network
with 3 input channels and 1 output channel, this time using the
compile-time API.
\begin{lstlisting}[label={lis:CompileTimeEx},captionpos=b,
    caption={\it Example of the RTNeural compile-time API}]
#include <RTNeural/RTNeural.h>
      
int main()
{
    // define model
    RTNeural::ModelT<double, 3, 1,
        RTNeural::DenseT<double, 3, 4>,
        RTNeural::TanhActivationT<double, 4>,
        RTNeural::DenseT<double, 4, 1>
    > model;

    // load model from json
    std::ifstream jsonStream("model.json", std::ifstream::binary);
    model.parseJson(jsonStream);
      
    model.reset(); // reset model state
      
    // set up input vector
    double input[] = { 1.0, 0.5, -0.1 };

    // compute output
    double output = model.forward(input);
      
    return 0; // exit
}
\end{lstlisting}

\section{Examples}
To date, \RT\ has mainly proven to be useful for real-time
audio processing tasks. Two real-world applications where
\RT\ is currently in use include guitar distorton effects,
and analog tape emulation.

\subsection{Guitar Distortion}
\cite{chowdhury:klon:2020} describes a real-time ``virtual
analog'' model of the famed Klon Centaur guitar distortion
circuit. While the effect was originally implemented using
a custom neural inferencing engine, more recent versions
of the effect use \RT\ instead. Switching from the custom
inferencing engine to \RT\ has resulted in a performance
improvement of nearly 3x. Further, the neural network
processing with \RT\ is approximately twice as fast as
an alternative model of the circuit, developed using more
traditional circuit modelling techniques.

\subsection{Analog Tape Emulation}
\cite{chowdhury:physModelTape:2019} derives a method for
creating a real-time digital emulation of a magnetic
tape recorder using physical modelling. In particular, the
hysteresis process that describes the magnetisation of
the tape is emulated using a discretized version
of the Jiles-Atherton equation \cite{JilesAtherton1986}.
While the method described in the paper uses the 4th-order
Runge-Kutta method to evaluate the equation in real-time,
more recent implementations use the 2nd-order Runge-Kutta
method, or a Newton-Raphson solver with either 4 or 8 maximum
iterations.
\newline\newline
Recently, the developer has introduced an alternative solver
based on a ``State Transition Network'' (STN) \cite {NLML},
using \RT\ for real-time inferencing. The trained STN achieves
similar accuracy to the 4-iteration Newton-Raphson solver
while performing nearly 2x faster. The process of training
and implementing the STN solver is decribed in \cite{tape_stn}.
\begin{figure*}[ht]
    \centering
    \includegraphics[width=0.33\textwidth]{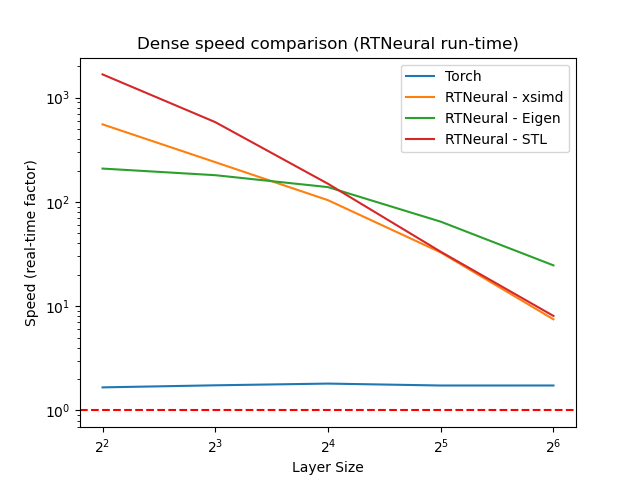}
    \includegraphics[width=0.33\textwidth]{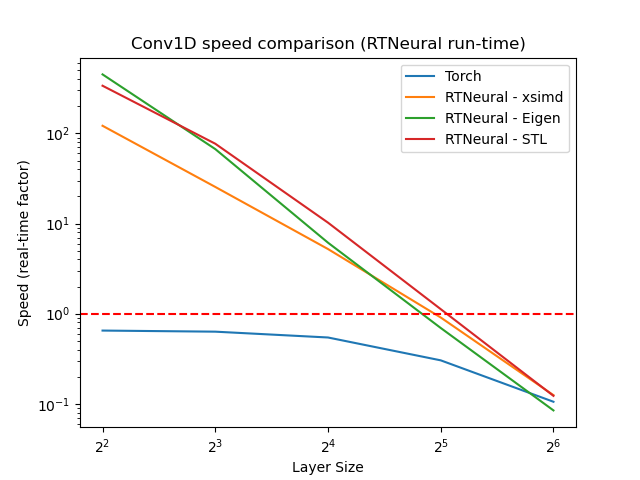}
    \includegraphics[width=0.33\textwidth]{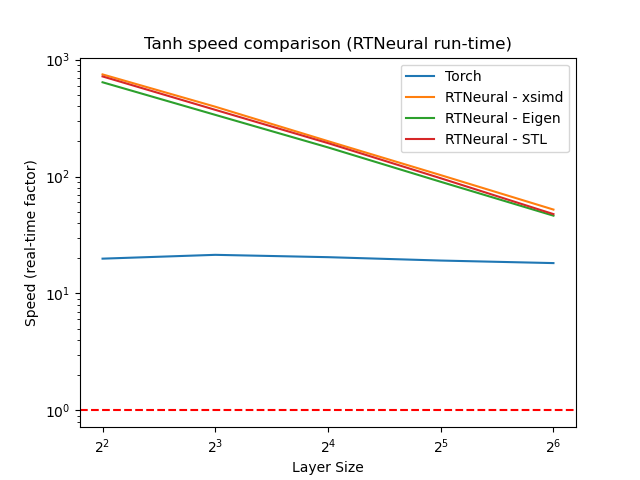}
    \caption{\label{fig:perf_dynamic} {\it Speed comparison for dense, Conv1D, and $\tanh$ layers, using the \RT\ run-time API.}}
\end{figure*}
\begin{figure*}[ht]
    \centering
    \includegraphics[width=0.33\textwidth]{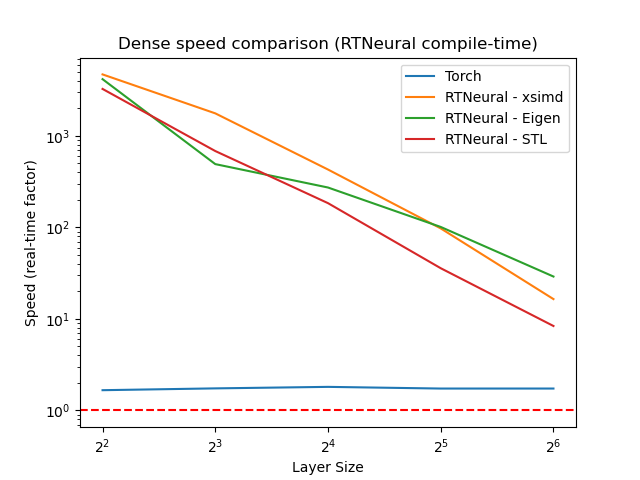}
    \includegraphics[width=0.33\textwidth]{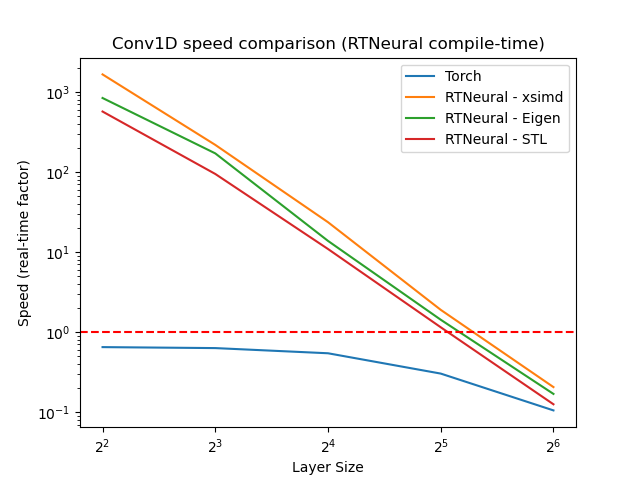}
    \includegraphics[width=0.33\textwidth]{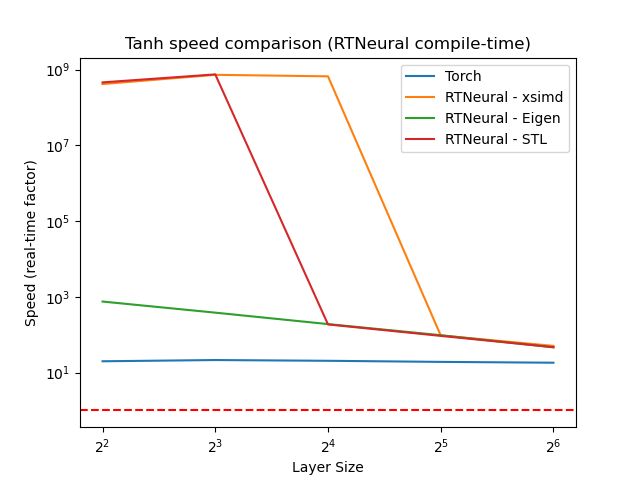}
    \caption{\label{fig:perf_static} {\it Speed comparison for dense, Conv1D, and $\tanh$ layers, using the \RT\ compile-time API.}}
\end{figure*}
\section{Performance Comparison}
In order to better understand the performance benefits of
\RT\ relative to other neural network inferencing libraries,
a performance benchmark has been developed comparing the speed
of each \RT\ backend against the PyTorch \CPP\ API. The benchmark
constructs a single layer with a given number of input and output
channels (referred to as the ``Layer Size'') and processes a random
signal at 48 kHz sample rate. The speed of the layer is then measured
as a ``real-time factor'' $v_{RT}$, defined as,
\begin{equation}
    v_{RT} = \frac{t_{signal}}{t_{process}}
\end{equation}
where $t_{signal}$ is the duration of the input
signal, and $t_{process}$ is the amount of time
taken by the layer to process the input. Note that
for $v_{RT} < 1$ the layer processing is too slow
to perform inferencing in real-time.
\newline\newline
\Cref{fig:perf_dynamic} shows the performance results
for a selection of layers using the \RT\ run-time API.
As a reference, the results are compared to the
performance of the PyTorch \CPP\ API, using the same
layers and layer sizes. Note that the dashed red line
in the plots marks the ``real-time'' threshold $v_{RT} = 1$.
\Cref{fig:perf_static} shows a similar set of results
for the \RT\ compile-time API. For the performance benchmarks
shown here, all measurements were made using a 2019 MacBook Pro
with an Intel\textregistered\ Core\texttrademark\ i7-9750H CPU,
running at 2.6 GHz.
\newline\newline
The results of the performance benchmarks show that
\RT\ out-performs PyTorch for the smaller layer sizes,
although PyTorch shows less change in performance
as the layer size grows, implying that it will be
better able to ``scale up'' for larger layer sizes.
This performance trade-off between larger and smaller layer
sizes is to be expected, since \RT\ is intentionally optimised
to perform better for smaller layer sizes. Comparing the
performance between the \RT\ backends shows that different
backends perform best for different layer types and sizes.
In general, it is recommended that users profile the performance
of the network they are implementing on their target hardware
to determine which backend is the best choice for their use-case.
\newline\newline
The source code for the performance benchmarks can be found
on GitHub,\footlink{https://github.com/jatinchowdhury18/RTNeural-compare}
along with a full set of plots displaying performance results
for all the neural layers supported by \RT.

\section{Conclusion}
\RT\ is a neural network inferencing library written in \CPP,
and is optimised for use in systems with hard real-time constraints.
The library is designed to be lightweight, by supporting only the
most relevant subset of neural network layers; flexible, by
supporting several computational backends and both run-time
and compile-time API's; and, most importantly, highly performant.
\newline\newline
The library has already proven to be useful for real-time
audio processing tasks, including guitar distortion and analog
tape emulation. Further, performance benchmarks show that \RT\ can
achieve superior performance compared to the PyTorch \CPP\ API
for smaller network sizes.
\newline\newline
Future work will involve implementing more commonly used
neural network layers, such as max-pooling \cite{max_pool}
and batch normalization \cite{ioffe2015batch}. Additionally,
new use-cases will be explored, including more audio-related
tasks such as randomised overdrive \cite{steinmetz2020overdrive}
and wake-word detection \cite{wake_words}, as well as real-time
tasks in other domains.

\section{Acknowledgments}
The author would like to thank Pete Warden and the EE292 class at
Stanford University for their inspiration, as well as Christian
Steinmetz and Keith Bloemer for their educational conversations.
Thanks as well to the Center for Computer Research in Music and
Acoustics (CCRMA) for providing relevant computing resources.

\bibliographystyle{IEEEbib}
\bibliography{references}

\end{document}